\documentclass{JINST}
\pdfoutput=1
\usepackage[pdftex]{graphicx}
\graphicspath{{images/}}
\usepackage{amsmath,amssymb}
\usepackage{array}
\newcommand{\plotc}{\includegraphics[clip=true,trim=1.2cm 0.5cm 1.5cm 1cm,width=0.2\textwidth]}

\usepackage{natbib}

\def\instIASFMI{$^{1}$}
\def\instUNIMI{$^{2}$}
\def\instLABEN{$^{3}$}
\def\instIASFBO{$^{4}$}
\def\instIFP{$^{5}$}
\def\instCANT{$^{6}$}
\def\instJBO{$^{7}$}
\def\instOATS{$^{8}$}
\def\instDA{$^{9}$}
\def\instYLI{$^{10}$}
\def\instESA{$^{11}$}
\def\instMILLI{$^{12}$}

\title{{\sc Planck}-LFI radiometers' spectral response}

\author{
A.~Zonca{\instIASFMI} \thanks{Corresponding Author, e--mail: andrea.zonca@fisica.unimi.it}
  ,
      C.~Franceschet{\instUNIMI}
  ,
      P.~Battaglia{\instLABEN}
  ,
      F.~Villa{\instIASFBO}
  ,
      A.~Mennella{\instUNIMI}
  ,
      O.~D'Arcangelo{\instIFP}
  ,
      R.~Silvestri{\instLABEN}
  ,
      M.~Bersanelli{\instUNIMI}
  ,
      E.~Artal{\instCANT}
  ,
      R.C.~Butler{\instIASFBO}
  ,
      F.~Cuttaia{\instIASFBO}
  ,
      R.J.~Davis{\instJBO}
  ,
      S.~Galeotta{\instOATS}
  ,
      N.~Hughes{\instDA}
  ,
      P.~Jukkala{\instDA}
  ,
      V.-H.~Kilpi\"{a}{\instDA}
  ,
      M.~Laaninen{\instYLI}
  ,
      N.~Mandolesi{\instIASFBO}
  ,
      M.~Maris{\instOATS}
  ,
      L.~Mendes{\instESA}
  ,
      M.~Sandri{\instIASFBO}
  ,
      L.~Terenzi{\instIASFBO}
  ,
     J.~Tuovinen{\instMILLI}
  ,
     J.~Varis{\instMILLI}
  ,
     A.~Wilkinson{\instJBO}\\
 \llap{\instIASFMI} INAF-IASF Milano,\\
 Via E.~Bassini 15, I-20133 Milano, Italy\\
 \llap{\instUNIMI}
 Universit\'a di Milano, Dipartimento di Fisica,\\
 Via G.~Celoria 16, I-20133 Milano, Italy\\
\llap{\instLABEN}
  Thales Alenia Space Italia S.p.A., \\
 S.S. Padana Superiore 290, 20090 Vimodrone (Mi), Italy \\
\llap{\instIASFBO}
 INAF-IASF Bologna, \\
     Via P.~Gobetti, 101, I-40129 Bologna, Italy\\
 \llap{\instIFP}
        IFP-CNR \\
        via Cozzi 53, 20125 Milano\\
 \llap{\instCANT}
	      Departamento de Ingenier\'{\i}a de Comunicaciones\\
          Universidad de Cantabria, Avenida de los Castros s/n. 39005 Santander, Spain\\
 \llap{\instJBO}
	      Jodrell Bank Centre for Astrophysics\\
          Alan Turing Building, The University of Manchester, Manchester, M13 9PL, UK \\
 \llap{\instOATS} INAF-OATs,\\
 Via G.B.~Tiepolo 11, I-34131, Trieste, Italy\\
 \llap{\instDA}
	      DA-Design Oy\\
          Jokioinen, Finland\\
 \llap{\instYLI}
	      Ylinen Electronics Oy\\
          Kauniainen, Finland\\
 \llap{\instESA}
              ESA - ESAC\\
              Camino bajo del Castillo, s/n, Villanueva de la Ca\~{n}ada 28692 Madrid\\
 \llap{\instMILLI}
	      MilliLab\\
          VTT Technical Research Centre of Finland, Espoo, Finland\\
	  }

 \abstract{
     The Low Frequency Instrument (LFI) is an array of pseudo-correlation radiometers on board the  {{\sc Planck}} satellite, the ESA mission dedicated to precision measurements of the Cosmic Microwave Background. The LFI covers three bands centred at 30, 44 and 70 GHz, with a goal bandwidth of 20\% of the central frequency. The characterization of the broadband frequency response of each radiometer is necessary to understand and correct for systematic effects, particularly those related to foreground residuals and polarization measurements. 
     
     In this paper we present the measured band shape of all the LFI channels and discuss the methods adopted for their estimation. The spectral characterization of each radiometer was obtained by combining the measured spectral response of individual units through a dedicated RF model of the LFI receiver scheme. As a consistency check, we also attempted end-to-end spectral measurements of the integrated radiometer chain in a cryogenic chamber. However, due to systematic effects in the measurement setup, only qualitative results were obtained from these tests. The measured LFI bandpasses exhibit a moderate level of ripple, compatible with the instrument scientific requirements.
}


\begin{document}

\section{Introduction}
\label{sec:intro}
The objective of the {\sc Planck} mission is to provide a definitive characterisation of CMB temperature anisotropies, a good detection of E mode polarisation and possibly measure B mode polarisation for the first time, see \cite{2009_COM_Mission}.
{\sc Planck} is equipped with two instruments: the HFI is a bolometric array covering frequencies between 100 and 857 GHz \cite{2009_HFI_Instrument}, while the LFI is a radiometric array covering the range of frequencies between 30 and 70 GHz, see \cite{2009_LFI_cal_M2}.

  \begin{figure}
    \centering
    \includegraphics[width=.9\textwidth]{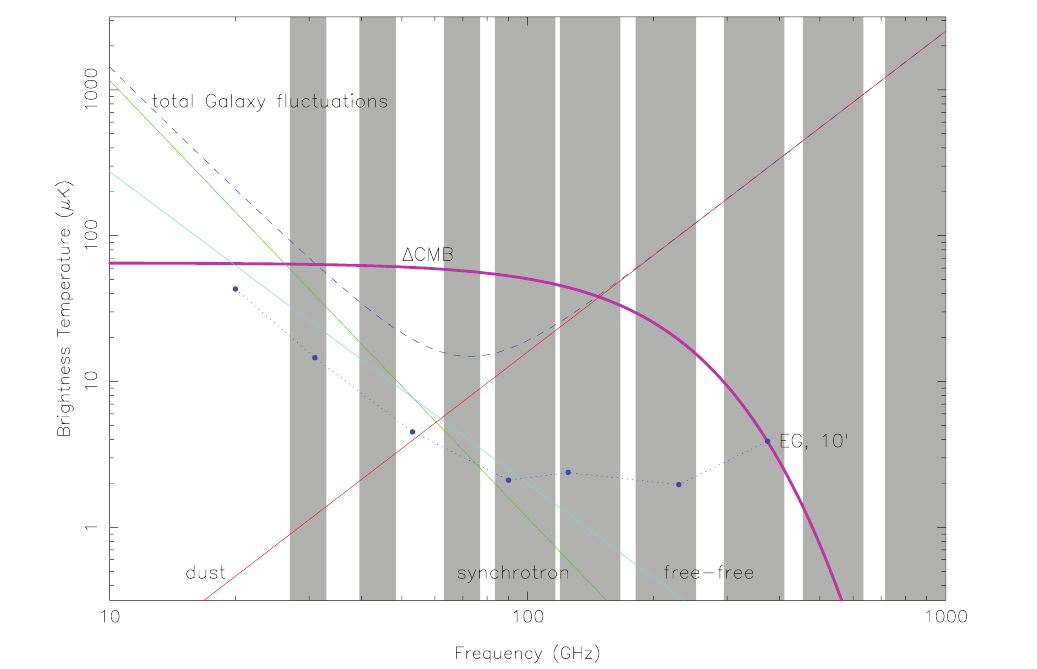}
    \label{fig:foregrounds}
    \caption{The spectra of synchrotron, dust, and free-free foregrounds superimposed on the {\sc Planck} channels' bandwidths and the expected CMB signal at 1 degree scale (image from \cite{planckbluebook})}
  \end{figure}

Accurate measurements of the CMB fluctuations require a robust detection and removal of foreground emission. Data analysis strategies rely on their different frequency dependence in order to distinguish them from the CMB.
Figure \ref{fig:foregrounds} shows the frequency dependence of dust, synchrotron, free-free diffuse emissions and the contribution from extragalactic sources, compared to the CMB. Thanks to its wide frequency coverage, {\sc Planck} will be able to separate with high precision the different components by combining maps from different channels.

Both radiometers and bolometers have a broadband response and their output is a convolution of the input spectra, weighted by the detector beam pattern, with their own frequency response. Good knowledge of the instrument's frequency response allows, if necessary, a fine correction to be applied at map level by an iterative process involving separation of the different components using their spectral indices, convolution of each foreground with the instrument frequency response and comparison of their sum with the observed map.

Bandpass knowledge is also important for CMB polarisation measurements, the new frontier of precision cosmology that {\sc Planck} is expected to open up.
The HFI has polarisation capability at 100, 143, 217 and 353 GHz thanks to Polarisation Sensitive Bolometers\cite{2009_HFI_Polarization} (PSB) and all LFI channels are polarisation sensitive.
Each LFI feed horn \cite{2009_LFI_cal_O1} is connected to an orthomode transducer \cite{2009_LFI_cal_O2} 
which splits the signal into two orthogonal linear polarisation components detected by two independent radiometers.
Each radiometer is calibrated separately against the CMB dipole signal and polarisation measurements are performed combining the signal received by pairs of feed horns located symmetrically in the focal plane and following each other in the scan circle. For a complete description of the LFI focal plane see \cite{2009_LFI_cal_M2}.
The main issue is that foregrounds have different spectral indices with respect to the CMB dipole which determined the calibration, therefore a spurious polarisation signal is generated when differencing channels with different bandpasses, producing leakage of the foregrounds' total intensity into the polarised components.  
Using the radiometers' frequency response, it is possible to estimate the contamination of bandpass mismatch on CMB maps, as described in \cite{2009_LFI_polarisation_M6}.

During qualification and performance tests, each active and passive unit composing LFI radiometers was characterised in terms of frequency response. For each unit we have measured return loss, insertion loss and (when applicable) gain and noise temperature as a function of frequency over the entire bandpass. These hardware measurements are combined together by the LFI Advanced RF model, developed since 2001 \cite{thesis_battaglia,thesis_franceschet}, which computes the bandpass of each channel of the integrated radiometer taking into account multiple reflections. For a complete review see \cite{2009_LFI_cal_R5}).

During the Flight Model (FM) test campaigns we attempted an end-to-end measurement of the radiometers' bandpasses on the integrated radiometer chains (see section \ref{sec:swept}), but this suffered from systematic effects in the measurement set-up.  
We have nevertheless been able to draw some qualitative conclusions from them.


\section{LFI radiometers' frequency response: basic equations and definitions}
\label{sec:freqresponse}

The LFI gathers photons from the sky through 11 corrugated feed horns. Each horn is connected to an orthomode transducer (OMT) which splits the incoming signal into two orthogonally polarised components that propagate through two independent pseudo-correlation receivers that continuously compare the sky signal with a stable reference load at the temperature of $\sim$4~K (\cite{2009_LFI_cal_R1}).

\begin{figure}[ht]
    \centering
    \begin{center}
    \includegraphics[width=10cm]{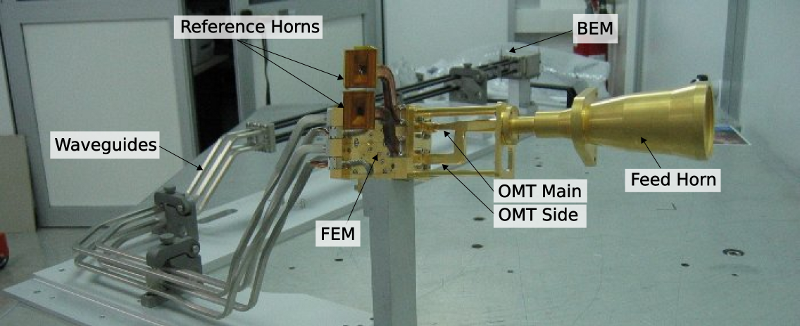}
\end{center}
    \caption{The 30 GHz Radiometer Chain Assembly (RCA) mounted on the support used during the cryogenic test campaign on the individual receivers.}
    \label{fig:rca_pic}
\end{figure}

Each of the 22 receivers has cryogenic front-end module (FEM at $\sim$20~K) and warm back-end module (BEM at $\sim$300~K) radio frequency (RF) amplification stages (\cite{2009_LFI_cal_R8,2009_LFI_cal_R9,2009_LFI_cal_R10}) connected by two composite waveguides \cite{2009_LFI_cal_O3} and two square-law detectors that convert the RF power into a DC signal (figure~\ref{fig:rca_pic}).

In section~\ref{sec:radresp} we describe the broadband response of the LFI radiometers. In section~\ref{sec:naming} we shall briefly present the naming conventions used to identify the parts of the radiometers.

\subsection{Radiometer response}
\label{sec:radresp}

Under the assumption of receiver linearity, the voltage output from each of the 44 output diodes can be represented mathematically by this equation:

\begin{equation}
   \bar{V}_{out} = \bar{G} \cdot \bar{P}_{in} + \bar{V}_{noise},
   \label{eq:v_out}
\end{equation}
where $\bar{P}_{in}$ represents the input RF power, $\bar{G}$ is the photometric calibration constant and $\bar{V}_{noise}$ is the noise contribution from the RF amplifiers and back-end electronics.

    It is common to refer this noise contribution to the input, so that: 

\begin{equation}
   \bar{V}_{out} = \bar{G} \cdot (\bar{P}_{in} + \bar{P}_{noise}),
   \label{eq:v_out_2}
\end{equation}
where $\bar{P}_{noise} = \bar{V}_{noise} / \bar{G}$. 

The power term $\bar{P}_{noise}$ can be expressed in terms of noise temperature $\bar{T}_{noise}$:

\begin{equation}
   \bar{T}_{noise} = \dfrac{\bar{P}_{noise}}{k_B \Delta \nu},
   \label{eq:t_noise}
\end{equation}
where $k_B$ is Boltzmann's constant, $ \Delta \nu $ is the bandwidth, $\bar{T}_{noise}$ is the input load temperature that would produce an output signal $\bar{V}_{noise}$ in an ideal system, i.e. a system with no noise with $\bar{V}_{out} = \bar{G} \cdot \bar{P}_{in}$.

The noise temperature is directly related to the radiometer sensitivity by the radiometer equation:
    
\begin{equation}
   \sigma_N = K \dfrac{\bar{T}_{in}+\bar{T}_{noise}}{\sqrt{\Delta \nu \cdot \tau}},
   \label{eq:radiometer_equation} 
\end{equation}
where:
\begin{itemize}
    \item  $\bar{T}_{in}$ is the Rayleigh-Jeans  brightness temperature of the input source,
    \item  $\sigma_N$ is the root mean square of the white noise,
    \item $K$ is a factor which is dependent on receiver architecture and for the LFI receivers $K=\sqrt{2}$ \cite{2009_LFI_cal_M2},
    \item $\tau$ is the integration time.
\end{itemize}
 
If we explicit the frequency dependence, Eq.~\eqref{eq:v_out} reads: 

\begin{equation}
V_{out}(\nu) = G(\nu) \cdot P_{in}(\nu) + V_{noise}(\nu),
\label{eq:v_out_3}
\end{equation}
where the function $G(\nu)$, i.e. the effective gain versus frequency, is also referred to as the receiver bandpass.
The relationship between integrated and frequency dependent quantities are described by the following equations:
\begin{eqnarray}
    \bar{V}_{noise} & = & \int_{0}^{\infty} V_{noise}(\nu) d\nu, \\
    \bar{P}_{in} & = & \int_{0}^{\infty} P_{in}(\nu) d\nu, \\
    \bar{G} & = & \dfrac{\int_{0}^{\infty} G (\nu) \cdot P_{in}(\nu) d\nu }{\bar{P}_{in}}.
    \label{eq:all_integrated} 
\end{eqnarray}

During the {\sc Planck} full-sky survey, the photometric calibration constant $\bar{G}$ will be routinely calculated from the modulation of the CMB dipole across the sky and then used to produce temperature maps\cite{calibration}.

A priori knowledge of the bandshape $G(\nu)$ is important for a detailed understanding of the data, particularly to ensure control of potential systematic effects that may be introduced by the presence of foreground components both in the temperature and in the polarisation CMB maps.
As evident from equation \eqref{eq:all_integrated}, calibration is tightly bound to the spectral index, i.e. frequency dependence of the input source.
CMB, free-free, dust and synchrotron foregrounds have different spectral indices and different couplings with each radiometric channel bandpass and in principle each would have a different photometric constant. For this reason, using the photometric constant derived from the CMB dipole introduces a second order systematic effect on these components.
Knowledge of the instrument bandpasses can be used to reduce the effect by using an iterative process to separate different components \cite{2009_LFI_polarisation_M6}.

The receiver bandpass, $G(\nu)$, can be determined by performing two measurements per each frequency at different input power levels:

\begin{equation}
   G(\nu) = \dfrac{\Delta V_{out}(\nu)}{\Delta P_{in}(\nu)}.
   \label{eq:bandpass_calculation}
\end{equation}

\subsection{LFI naming convention}
\label{sec:naming}

Each of the 11 assemblies connected to a feed horn is called a Radiometer Chain Assembly (RCA) and is identified with a number ranging from 18 to 28 according to the following scheme:
\begin{itemize}
    \item RCA numbers 18 through 23: 70 GHz,
    \item RCA numbers 24 through 26: 44 GHz,
    \item RCA numbers 27-28: 30 GHz.
\end{itemize}

Each of the two radiometers connected to an RCA is labelled either with \texttt{M}
is labeled either with \texttt{M} or \texttt{S}, whether it is connected to the main arm or the side arm of the orthomode transducer (OMT).
Finally the four detector output diodes of each RCA are labelled \texttt{00} and \texttt{01} if they are connected to \texttt{M} and \texttt{10} and \texttt{11} if they are connected to \texttt{S}. Therefore the label \texttt{LFI25M-01} indicates the detector \texttt{01} in the radiometer \texttt{M}, i.e. connected to the OMT main arm of the  44 GHz RCA LFI25.

For a detailed review of the LFI instrument refer to \cite{2009_LFI_cal_M2}.

\section{Bandpass estimation}
\label{sec:bandpass_simulations}

Our most accurate method to measure the LFI bandpasses is based on measurements of individual components integrated into the LFI Advanced RF Model (LARFM) to yield a synthesised radiometer bandpass.

The LARFM is a software tool based on the Advanced Design System (ADS) circuit simulator by Agilent\footnote{Agilent Technologies, Inc http://www.home.agilent.com/agilent/product.jspx?nid=-34346.0.00}. 
The measured frequency responses of the various subsystems (feed-OMT, FEM, BEM) are considered as lumped S-parameter components.

Waveguides are simulated with an analytical model, in order to reproduce the effect of their temperature gradient\footnote{Recall that the LFI waveguides connect the 300~K BEM to the 20~K FEM} and the effect of standing waves caused by impedance mismatch at the interfaces between the FEM and BEM.
This is because the 1.8-meter long waveguides were not measured at unit level in cryogenic conditions. The model provides accurate agreement with the measured waveguide response in the conditions of the test measurements (300\,K).

\subsection{Model implementation}

\begin{figure*}
    \centering
    \includegraphics[width=\textwidth]{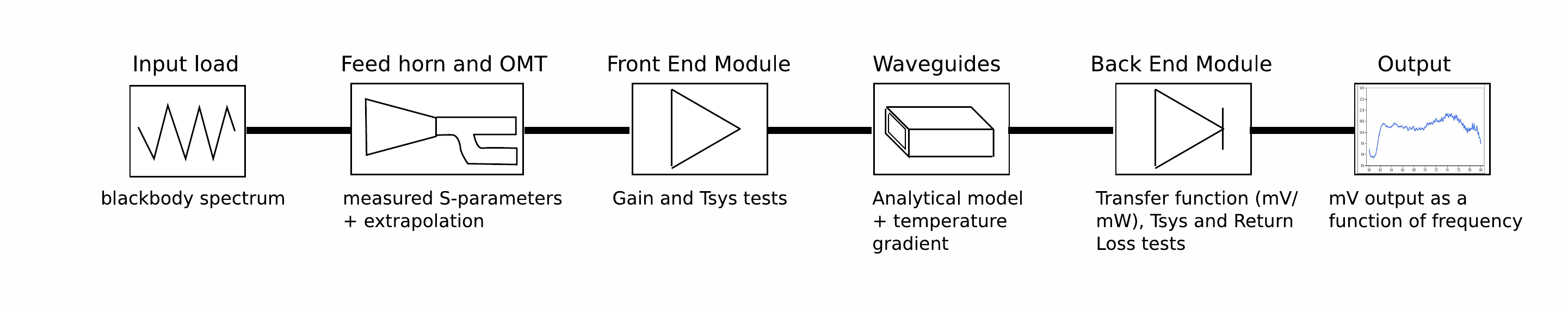}
    \caption{Model schematic for each LFI channel: 
    feed horn and OMT are characterized by their S-parameters; FEM by gain and noise temperature; waveguides are simulated analytically; BEM by return loss, transfer function and integrated noise temperature.}
    \label{fig:model_schematic}
\end{figure*}

The LARFM reproduces the frequency response of the 44 channels when looking at the input load. The overall layout of the model is sketched in figure~\ref{fig:model_schematic}. In this section we discuss the key points of each part that are relevant for bandpass response characterization. The detailed implementation can be found in \cite{2009_LFI_cal_R5}.  
The frequency dependence of all parameters is implicitly assumed if not otherwise stated.

\subsubsection{Input load}

The input load is used to provide different input power in order to compute the spectral response.
It is simply modelled as a resistor with tunable temperature; its output is an uncorrelated noise signal independent of frequency whose power spectral density is:

\begin{equation}
P_{in} = k_B T_{in} \quad [W/Hz].
\label{eq:sky_load}
\end{equation}

\subsubsection{Feed horn and orthomode transducer} 
\label{sec:mod-omt}

The feed horn and orthomode transducer are modeled as a single lumped two-port S-parameters component, characterized by scattering data taken at ambient temperature, see \cite{2009_LFI_cal_O2}. Reflection scattering parameter (S11 and S22) measurements were performed on the sub-assembly comprising feed horn and OMT. Return losses were obtained by injecting a signal into the rectangular port facing the FEM. For the insertion scattering parameters (S21 and S12), the results of the OMT tests are representative of the whole assembly, feed horn losses being very low (typically $< 1$ dB).

The 30 GHz OMT performance was measured between 26.5 and 40 GHz. In this case, all S parameters have been extrapolated down to 21 GHz, based on the 44 GHz OMTs which were measured over a very large band. The extrapolation was based on the normalization of each measurement at its own central frequency.

It should be noted that our calculation of the radiometers' frequency response does not include the frequency dependence of the optical system, i.e., of the feed-telescope beam pattern. This has been discussed in detail by \cite{phd_thesis_sandri,2009_LFI_cal_M5}. A full account of the instrument frequency response will require the combination of the optical and radiometric spectral dependences.

\subsubsection{Front End Module} 
\label{sec:mod-fem}
The performance of the FEMs was measured during validation tests performed by the FEMs' producers: Jodrell Bank Observatory (JBO) for 30 and 44 GHz devices\cite{2009_LFI_cal_R8}, and Ylinen for 70 GHz\cite{2009_LFI_cal_R10}.
Data are available both for FEM subcomponents, i.e. low noise amplifiers, hybrids and phase shifters, and for the integrated FEMs. The performance of sub components is strongly influenced by connections and by mutual interaction, therefore it is more reliable to consider tests of the complete FEM.

The FEMs' response is very linear and can be represented accurately by a linear model:
\begin{equation}
 P_{outFEM}(\nu) = G_{FEM}(\nu)\, k_B\, ( T_{in}(\nu) + T_{noiseFEM}(\nu) ),
\end{equation}
where:
\begin{itemize}
    \item $P_{outFEM}(\nu)$ is the output power per unit bandwidth;
    \item $G_{FEM}(\nu)$ is the dimensionless gain factor;
    \item $T_{noiseFEM}$ is noise temperature, mainly due to the FEM HEMT amplifiers \cite{2009_LFI_cal_R8}.
\end{itemize}

Gain and noise temperatures were computed using the Y-factor method, which consists of fitting the linear model presented above with measurements at two different temperatures:

\begin{align}
G_{FEM}(\nu) & = \dfrac{P_{outThigh}(\nu)-P_{outTlow}(\nu)}{k_B(T_{high}(\nu)-T_{low}(\nu))}, \\
Y(\nu) & = \frac{P_{outThigh}(\nu)}{P_{outTlow}(\nu)}, \\
T_{noiseFEM}(\nu) & = \dfrac{T_{high}(\nu) - Y(\nu) T_{low}(\nu)}{Y(\nu) - 1}.
\end{align}

The front-end modules are modelled as amplifiers characterized by a spectral dependence of gain and noise temperature as measured in the FEM units. The measurements were obtained with the Y-factor method at the cryogenic temperature for each LFI radiometer FEM channel. The return loss parameters S11 and S22 were inferred from the mean return loss measured at ambient temperature.

All measurements covered the full radiometer bands. However, at 30 GHz, the out-of-band low frequency tail from 25 to 21.3 GHz, the WR28 waveguide cutoff, could not be measured directly. In this case we extrapolated the out-of-band profiles using a set of measurements performed on a larger frequency span of one of the FEMs with a Vector Network Analyser (VNA). 

\subsubsection{Waveguides} 
\label{sec:mod-wg}

The 44 LFI composite waveguides are 1.5 to 1.8 meters long, depending on the channel, and they have a complex routing to allow integration of the HFI in the central portion of the focal plane. Furthermore, the waveguides support the thermal gradient between the $\sim20$\,K front end and the $\sim300$\,K back end, with three intercepts at the three V-grooves (see \cite{2009_COM_Mission}). This complex thermo-mechanical layout led to the choice of measuring the waveguide responses only at room temperature, and then the results were extrapolated to the results at nominal thermal conditions. This choice is justified by the fact that, while the cryogenic setup for a direct cryogenic measurement would be extremely complex and likely to be affected by systematic effects, the RF model of a waveguide is very well defined once its manufacturing characteristics are known.

We implemented a software simulator using several ADS internal rectangular waveguide components connected in cascade with temperatures linearly interpolated from  20 to 300\,K.

This model was compared successfully to ambient temperature measurements and it is able to reproduce correctly the S-parameters of the waveguides.

\subsubsection{Back End Module} 
\label{sec:mod-bem}

The BEM provides amplification (about 25 dB) and detection of the RF signal, and outputs a DC voltage that is integrated in time and digitalized by the LFI Data Acquisition Electronics.
Each of the four channels of a BEM contains an RF amplification stage based on HEMT Low Noise Amplifiers, a bandpass filter, a square law detector diode, and a DC amplifier. 

The overall frequency response of each BEM has been measured for each channel and is used as an input to the LARFM.
In particular, each channel is implemented with an S11 S-parameter component (just modelling the return loss), a noise source and a function which computes the incoming power and multiplies it by the BEM gain. All the return loss S-parameters, the transfer function (RF input power to Volt output gain) were directly measured at each BEM. The noise source temperature was derived by integrating the measured noise temperature.

A detailed description of the BEM devices and performance tests is available in \cite{2009_LFI_cal_R9} for 30 and 44 GHz and \cite{2009_LFI_cal_R10} for 70 GHz.

\subsection{Model output}

The LARFM computes the power due to thermal noise emitted by the input load at a given temperature in steps of 0.1 GHz, then propagates it through all RCA components considering multiple reflections. 
As discussed in the previous section, feed horn/OMT, FEM and BEM responses are defined by experimental data at each frequency, while the waveguides response is simulated by the software model itself.

To compute the bandpass, the simulation is run twice at different input load temperatures (2.7\,K and 200\,K) and equation~\eqref{eq:bandpass_calculation}  is used.
The concept of the procedure is similar to that of a swept source measurement, where a monochromatic power input is swept through the bandpass whilst recording the voltage output. 

As an independent check, we have carried out a swept source test of all RCAs in the cryogenic chamber.  Before presenting the bandpass results (section~\ref{sec:comparison}) we shall briefly present the end-to-end measurements in section \ref{sec:swept}.

\section{End-to-end cryogenic measurements}
\label{sec:swept}

\begin{figure}[h]
    \centering
    \includegraphics[width=.6\textwidth]{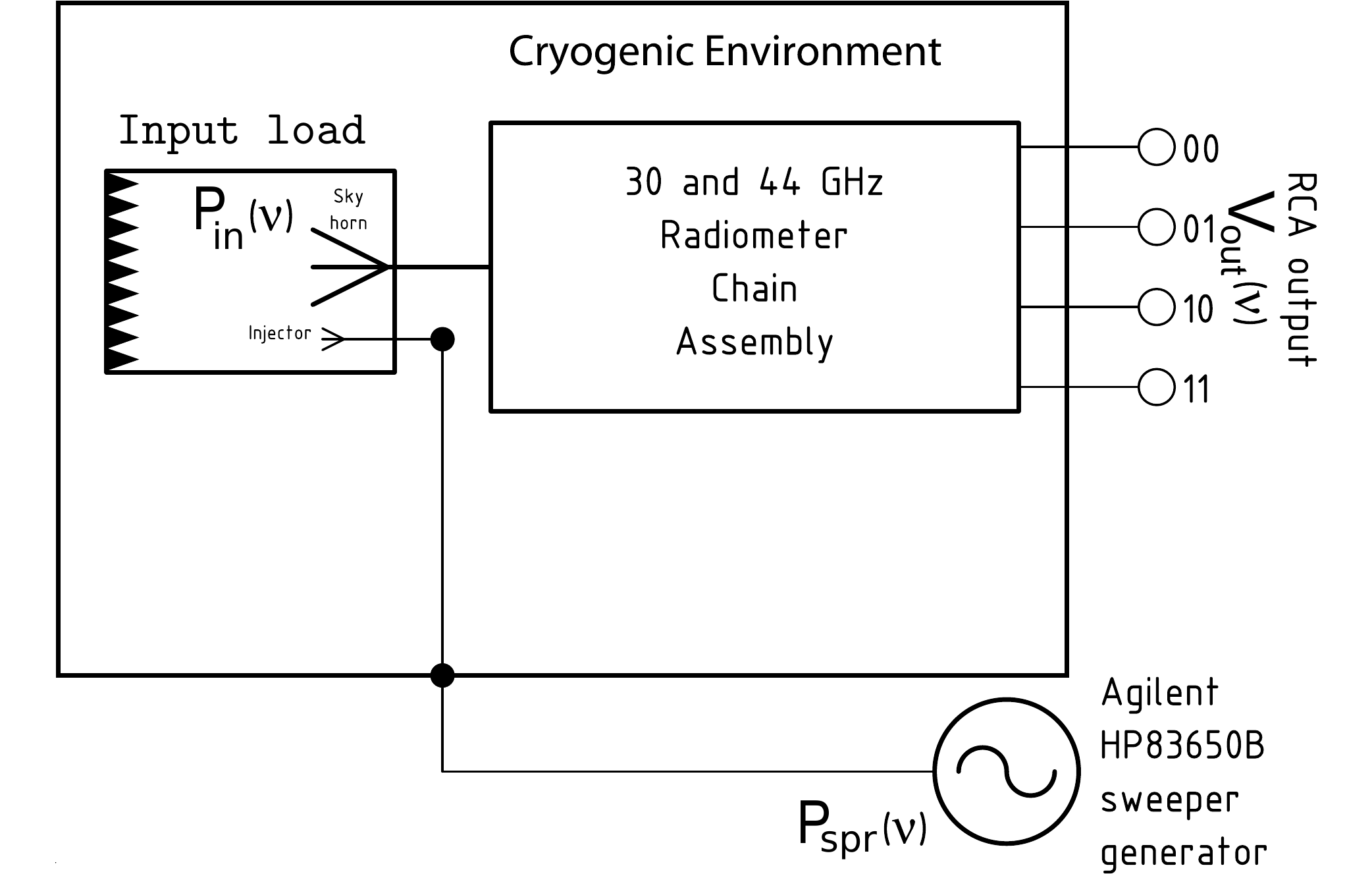}
    \caption[]{Schematic of the experimental setup for 30/44 GHz swept source tests: a monochromatic input microwave signal is injected into the input load. The input frequency changes with time across the radiometer band. The bandpass was computed from the recorded output voltage using Eq.~\eqref{eq:bandpass_calculation}.}
    \label{fig:sprsetup}
\end{figure}

The relative shape of the bandpass of each RCA was measured as part of the RCA test campaign by injecting a monochromatic signal into the input horn sweeping through the operational band and recording the DC output as a function of the input frequency, as shown in figure~\ref{fig:sprsetup}. 

\begin{figure}[h]
    \label{fig:lfi2711}
    \centering
    \includegraphics[width=.6\textwidth]{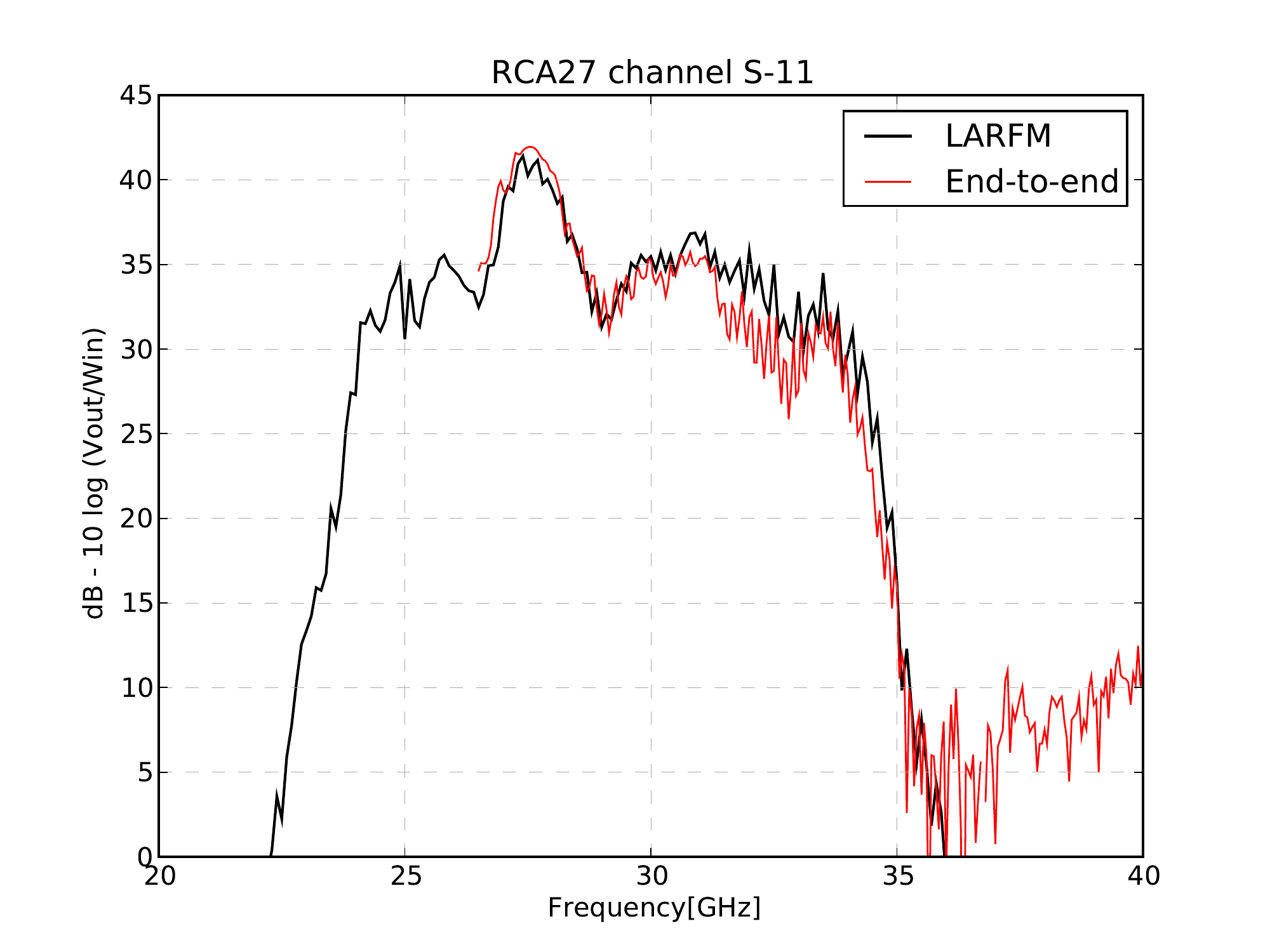}
    \caption{LFI27S-11 at 30 GHz. Comparison of LARFM bandpasses with end-to-end swept source measurements performed on WR28 nominal bandpass, between 26.5 and 40 GHz.}
\end{figure}

\begin{figure}[h]
    \label{fig:lfi2601}
    \centering
    \includegraphics[width=.6\textwidth]{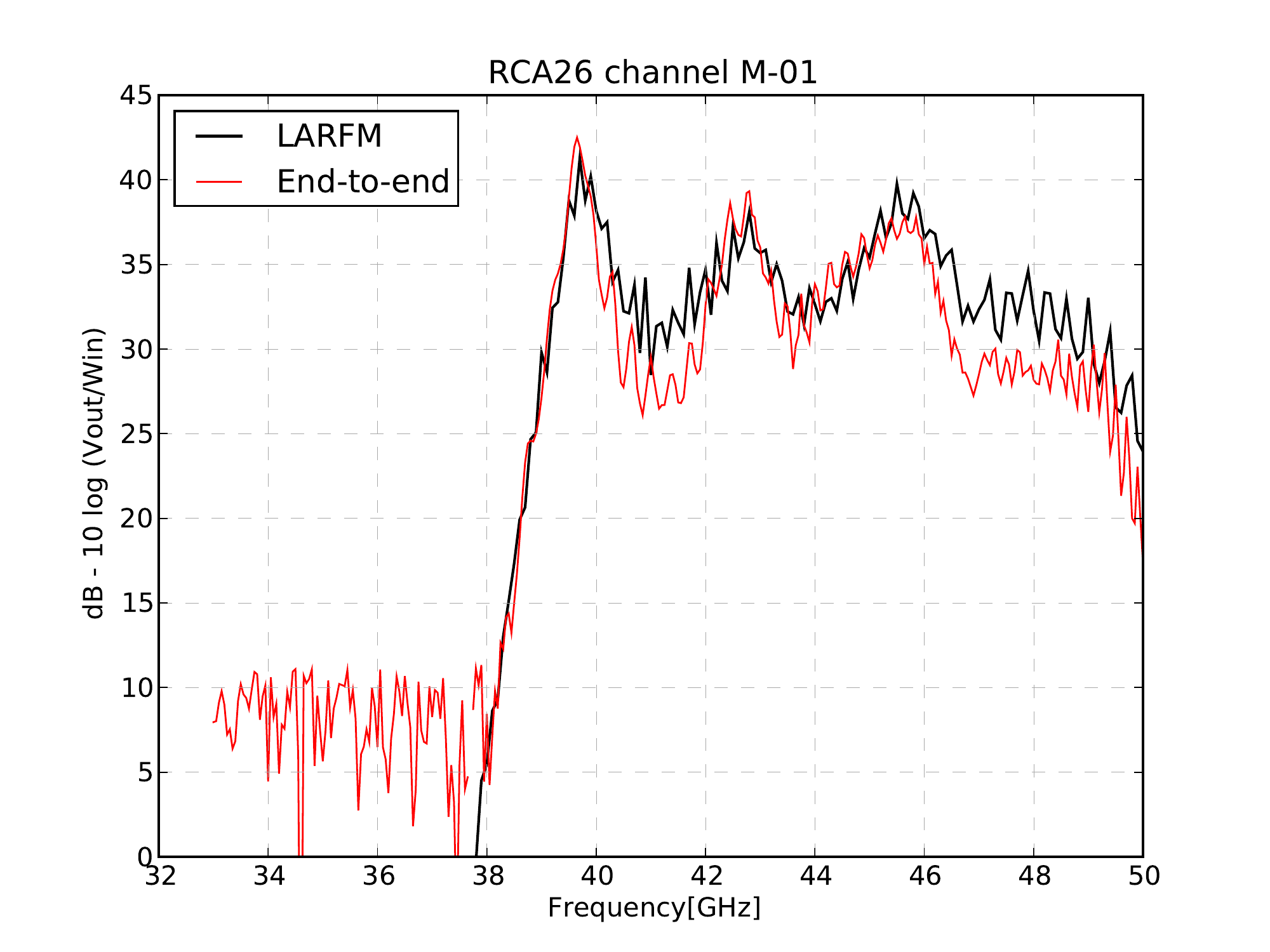}
    \caption{LFI26M-01 at 44 GHz. Comparison of LARFM bandpasses with end-to-end swept source measurements.}
\end{figure}

The 30 and 44 GHz RCAs were assembled and tested in Thales Alenia Space in Vimodrone, while the 70 GHz RCAs were characterised at DA-Design Ylinen in Finland. The experimental setup was slightly different in the two cases. At 30 and 44 GHz, the input load was a cylindrical blackbody cavity 20 cm high and with a diameter of 20 cm, providing a return loss of about 60 dB. The injector was an open-ended WR28/WR22 waveguide located in such a way that the signal was propagated into the RCA horn by the input load back wall. The results show, in general, good agreement with the bandpasses estimated by LARFM (see figures~\ref{fig:lfi2711} and \ref{fig:lfi2601}).

However, two different kinds of systematic effect hampered the measurements: in many 44 GHz channels a lack of power was observed in the high frequency portion of the spectrum; furthermore, in four cases, standing waves were produced as a consequence of poor alignment of the polarised RF injection into the input load.

At 70 GHz, the microwave generator was connected to a small input load placed directly in front of the horn, which yielded a return loss below the -30 dB requirement. This caused standing waves between the injecting horn and the RCA feed horn.

In conclusion, while the end-to-end tests yielded a useful qualitative cross-check, these tests did not produce reliable results for a quantitative analysis. The LFI bands are determined by the LARFM combined approach already discussed.

\section{Results}
\label{sec:all_plots}
\label{sec:comparison}

\begin{figure}[h]
    \label{fig:band70}
    \begin{center}
    \begin{tabular}{m{0.3cm} c c c c} 
      & \textbf{M-00} & \textbf{M-01} & \textbf{S-10} & \textbf{S-11}\\

\input{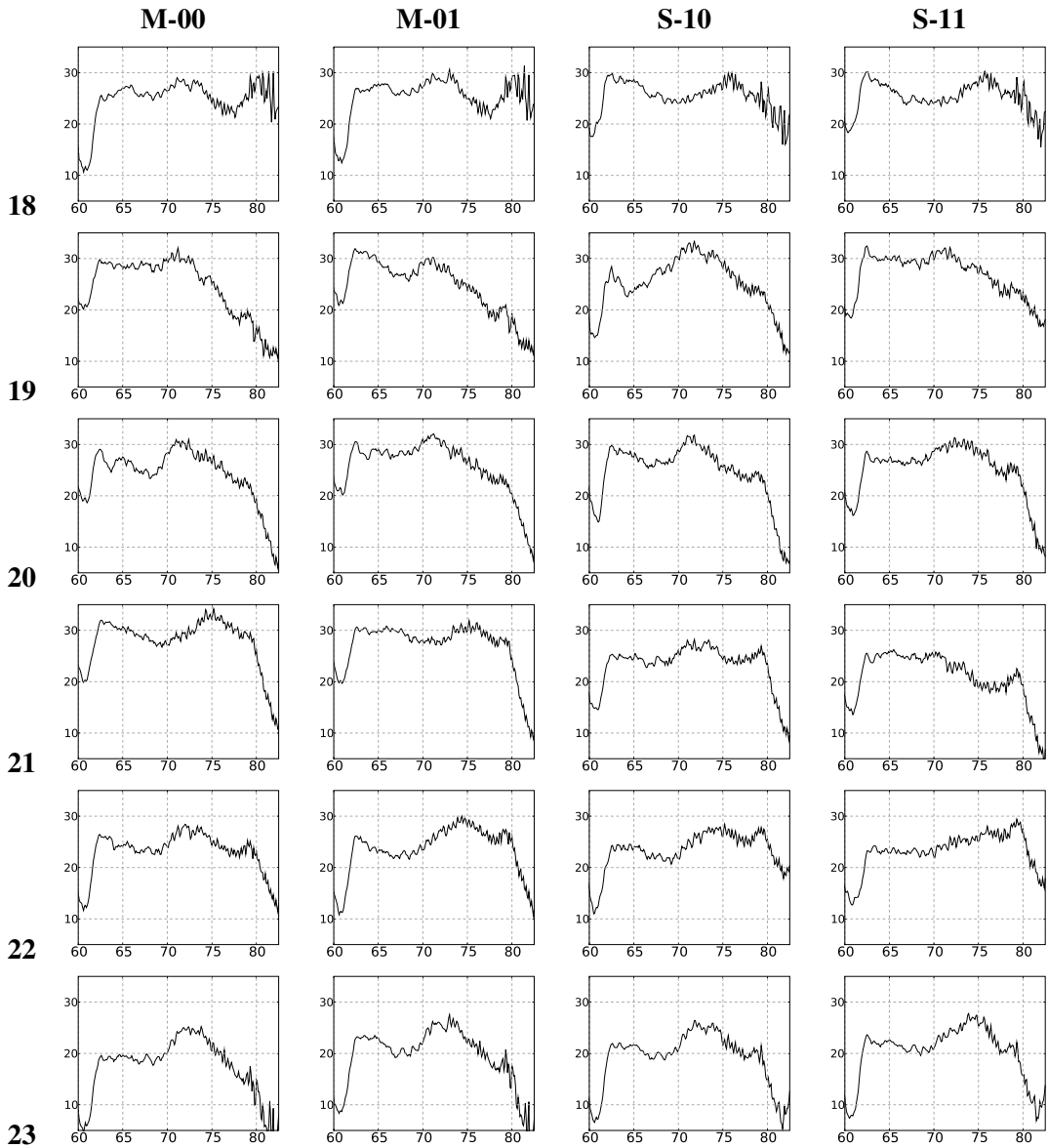}

    \end{tabular} 
    \caption[]{LFI 70 GHz channels' bandpasses. Each row shows the 4 bandpasses of a RCA ordered as \texttt{M-00}, \texttt{M-01}, \texttt{S-10} and \texttt{S-11}. Units are $[dB]$ $10 log (\frac{V_{out}}{V_{in}})$ plotted against frequency $[GHz]$.}
    \end{center}
\end{figure}

\begin{figure}[h]
    \label{fig:band3044}
    \begin{center}
    \begin{tabular}{m{0.3cm} c c c c} 
   & \textbf{M-00} & \textbf{M-01} & \textbf{S-10} & \textbf{S-11}\\

\input{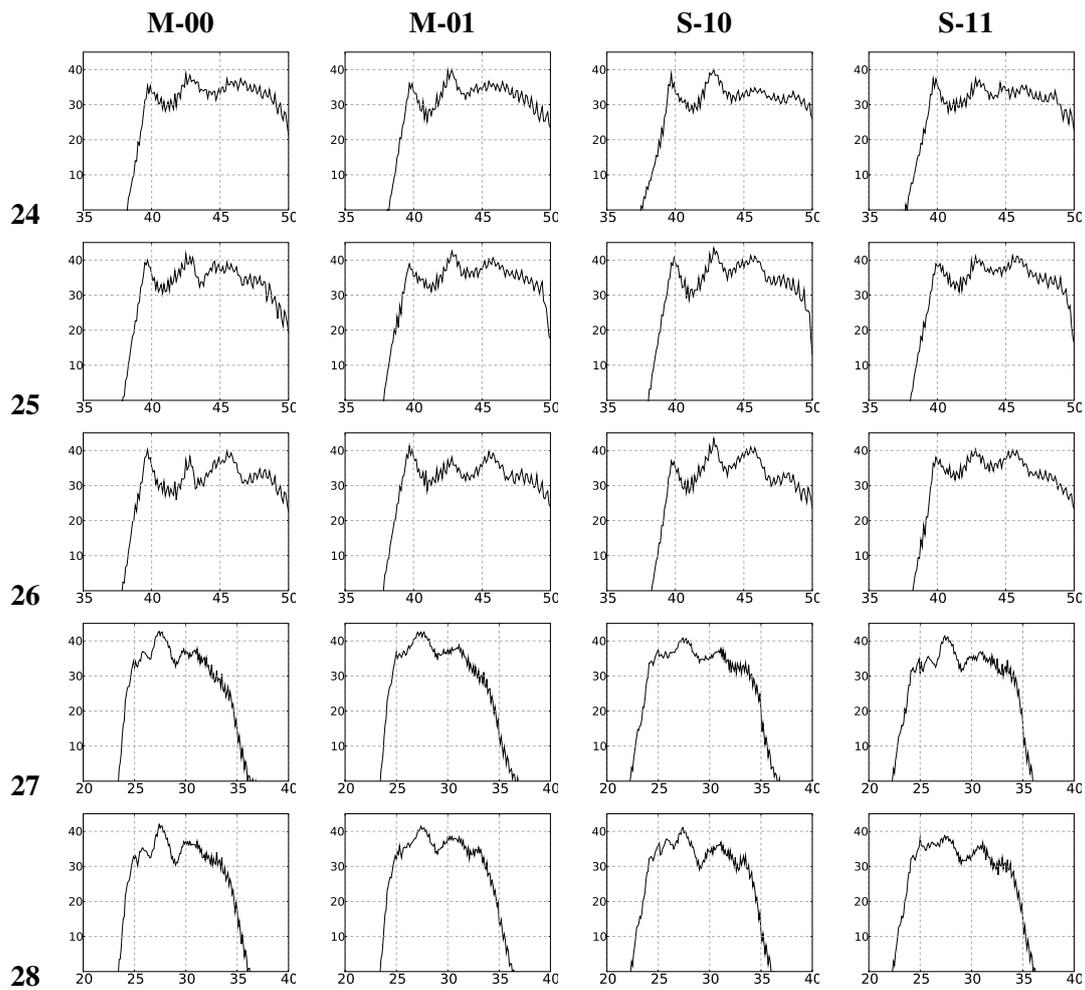}

    \end{tabular} 
    \caption[]{30 and 44 GHz bandpasses, formatted as figure \ref{fig:band70}} 
    \end{center}
\end{figure}

Figures \ref{fig:band70} and \ref{fig:band3044} show all the LFI bandpasses obtained by the frequency response data of each radiometer unit assembled by the LARFM.
The 70 GHz channels show a low bandpass ripple, of about  $10$ dB, which is within scientific requirements. The spike between 60 and 61 GHz, below the low frequency cut-off, is due to a systematic effect present in all the BEM gain measurements and caused by the test setup. We removed this range from the bandpasses made available at the Data Processing Center in order to avoid possible spurious effects and therefore the frequency coverage is 61-80 GHz. The high frequency cut-off is not well defined in most of the channels.

The 30 and 44 GHz bandpasses show a more complex shape, driven by the BEM spectral response, but still within $\pm10$ dB. The low frequency cut-off is always well defined, while the high frequency cut-off is not well defined in RCA 24 and 26. However, comparing with the high frequency cut-off of RCA 25, it is expected that the additional bandwidth is very low. Frequency coverage is 25-50 GHz for the 44 GHz channels and 21.3-40 GHz for the 30 GHz channels.


\clearpage

\newpage
\clearpage

\section{Conclusion}
\label{sec:conclusion}

We have presented the bandpasses of all the {\sc Planck}-LFI radiometers as computed using the LFI Advanced RF Model based on the measured response of single radiometer units. Measurements of single components (feed-horns, orthomode transducers, front-end modules, back-end modules) are obtained with standard methods, typically employing a VNA test set, and provide highly reliable results, with precision of order 0.1-0.2 dB over the entire band. For the waveguides we have used simulated inputs that closely match the measured data. The composite bandpasses are estimated to have a precision of about 1.5 to 2 dB. 

We also attempted an end-to-end measurement of the RCA spectral response in the cryofacility as an independent check. Unfortunately, these measurements suffered some subtle systematic effects in the test setup (standing waves at 70 GHz; polarisation mismatch and low frequency coverage at 30 and 44 GHz), preventing an accurate cross-check. However, the comparison shows a general agreement within limits of the test reliability and repeatability. 

\begin{figure*}
    \centering
    \includegraphics[width=\textwidth]{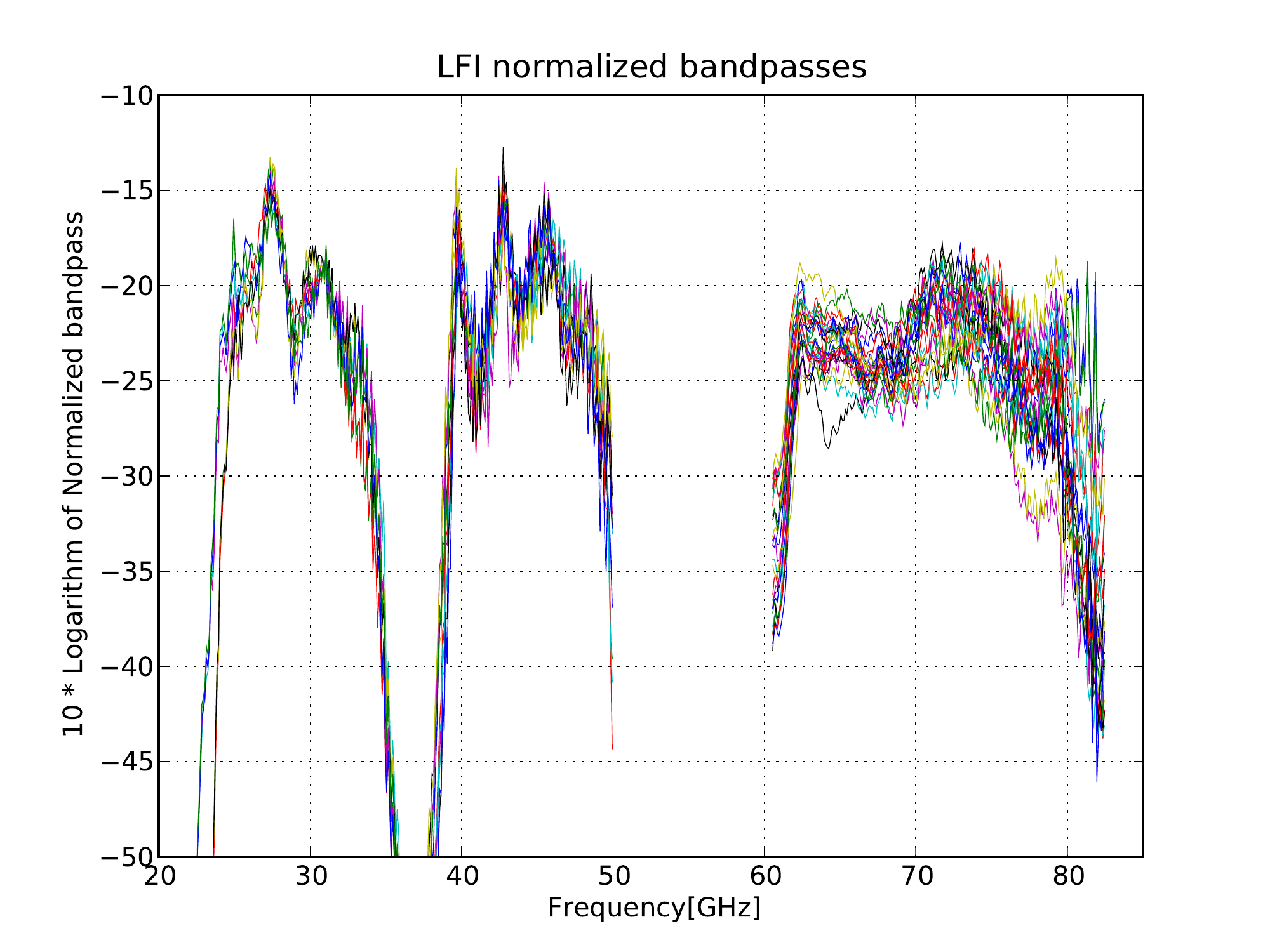}
    \caption{All the 44 LFI channels bandpasses made available to the LFI Data Processing Center in July 2009. Bandpasses are normalised to unit integral and plotted in logarithmic format computed as $10 \log w_i $ where each $w_i$ is the radiometer gain $G(\nu)$ integrated over 0.1 GHz, i.e. the bandpass sampling. In the release, 70GHz bandpasses below 60.5 GHz were removed because they were affected by a spike related to a systematic in the BEM measurements.}
    \label{fig:lfibands}
\end{figure*}

In figure~\ref{fig:lfibands}  we show the normalised bandpasses for all the 44 LFI channels. The spectral responses exhibit a moderate level of ripple, within $\pm 10$ dB, dominated by the response of the back-end modules at 30 and 44 GHz, which reduces the radiometer effective bandwidth. The measured band shapes are in line with the scientific needs of the instrument and the knowledge achieved in these measurements allow us to analyse the impact of the radiometers' frequency response on the scientific outputs.

The LFI Advanced RF Model is currently being ported to QUCS\footnote{http://qucs.sf.net}, an open source microwave circuits simulator. This development is expected to lead to a more flexible and linear implementation, see \cite{thesis_franceschet}, and to facilitate the interface with external scripts to run simulations and export results in view of the forthcoming {\sc Planck} data analysis.

\acknowledgments
  The work reported in this paper was carried out by the LFI instrument team of
  the {\sc Planck} Collaboration.
  {\sc Planck} is a project of the European Space Agency with instruments
funded by ESA member states, and with special contributions from Denmark
and NASA (USA). The {\sc Planck}-LFI project is developed by an International
Consortium lead by Italy and involving Canada, Finland, Germany, Norway,
Spain, Switzerland, UK, USA.

  The Italian group was supported in part by
  ASI (contract {\sc Planck} LFI Phase E2 Activity).

\bibliographystyle{plainnat}
\bibliography{aa_bandshape,references_prelaunch_forJI}

\end{document}